\documentclass[12pt]{article}
\usepackage{epsfig}

\def\simge{\mathrel{%
   \rlap{\raise 0.511ex \hbox{$>$}}{\lower 0.511ex \hbox{$\sim$}}}}
\def\simle{\mathrel{
   \rlap{\raise 0.511ex \hbox{$<$}}{\lower 0.511ex \hbox{$\sim$}}}}

\newcommand{\beq}{\begin{equation}}
\newcommand{\eeq}{\end{equation}}

\newcommand{\dd}{\partial}

\newcommand{\di}{\displaystyle}
\newcommand{\ga}{\gamma}
\newcommand{\scN}{{\scriptscriptstyle N}}
\newcommand{\scV}{{\scriptscriptstyle V}}

\newcommand{\si}{\sigma}
\newcommand{\ve}{\varepsilon}

\newcommand{\GeV}{\;{\rm GeV}}
\newcommand{\mkb}{\;{\rm \mu b}}

\title{Total $\gamma N$ cross section in the energy range
$\sqrt{s}=40-250$ GeV}

\author{O. Lalakulich${}^{1,2}$, \and Yu. Novoseltsev${}^{1}$,
\and R. Novoseltseva${}^{1}$, and  G. Vereshkov${}^{1,2}$}

\date{${}^{1}$ Institute for Nuclear Research of Russian Academy of Sciences,
117312 Moscow, Russia \\
${}^{2}$ Research Institute of Physics, Rostov State University, 344090
Rostov-on-Don, Russia}

\begin{document}

\begin{titlepage}

\maketitle

\begin{abstract}
The results of measurements of $\ga N$ total cross section, obtained by
the method of photoproduction processes registration at the Baksan
Underground Scintillation Telescope, are presented. These data at
energies $\sqrt{s}=40-130\GeV$ confirm the effect of more rapid
photon-hadron cross-section rise as compared to the hadron-hadron ones.
It is shown, the increasing of the additive quark number in the
products of photon hadronization can be one of the causes responsible
for this effect. On the basis of the analysis of experimental data on
both $\ga N$ and $\ga\ga$ total cross sections, the status of direct
and indirect cross-section measurements  is discussed.
\end{abstract}



\end{titlepage}

\section{Introduction. Status of the data.}

Since the first measurements from ZEUS \cite{DESYep} using 1995 data,
information about the proton (nucleon) structure function $F_2$ at low
$Q^2$ and $\ga N$ cross section $\si_{\ga N}(s)$ at high energies
generates a lot of interest. In the energy range $\sqrt{s}=100 \div 250$
GeV, the total cross section of $\ga p$ interaction were determined by
indirect method \cite{ekstrap} on the basis of DESY data on proton
structure functions \cite{DESYep}. At several points of energy scale
around $\sqrt{s}=200$ GeV, the cross section was measured by direct method
--- by photoproduction processes registration \cite{zeus94,aid95}. The
results obtained testify to more rapid photon--hadron cross section rise
in comparison with the hadron--hadron ones.
The physical nature of this
more rapid rise is the main problem of $\ga N$ physics, which is discussed
in many papers.  Modern interpretation in the framework of a Regge
parametrization is done in \cite{dl}. The reviews of experimental data
and theoretical approaches are given in \cite{amelung,godbole}.

In this paper we present the results of measurements of $\si_{\ga
N}(s)$ in the region $\sqrt{s}=40 \div 130$ GeV obtained at the Baksan
Underground Scintillation Telescope (BUST) of Baksan Neutrino
Observatory. In this experiment the hadronic and electromagnetic
cascades produced by cosmic ray muons are registered.
The BUST is situated at an effective depth of $8.5\times 10^4\ g/cm^2$
underground \cite{Alex79}. So only muons (with energy
$E_{\mu}>220\GeV$) and neutrinos can reach the telescope. That is
the experiment conditions correspond to the pure muon
beam with the known energy spectrum.
Preliminary results were presented earlier \cite{novo89}, however the data
on cross section $\si_{\ga N}(s)$ were not reported in \cite{novo89}.
The data obtained in this paper confirm the
effect of more rapid rise of photon-hadron cross section.

The theoretical basis of how to find the $\ga N$ cross section from the
data on muon interaction with nuclei was developed in \cite{bb},
\cite{bb2}. It was shown in this Refs., that in BUST experiment the number
of hadronic cascades is proportional to $\si_{\ga N}(s)$.

To interpret the experimental data, we propose a phenomenological theory
based on the Vector Dominance Model (VDM) and the Additive Quark Model
(AQM). In this theory the violation of photon--hadron scaling is
parametrised by two effects: 1) the rise of hadronization probability; 2)
the rise of hadronization product,  which is due to vector resonances
decay on pions. It is shown, the experimental data on $\ga N$ cross
section allows to find the measures of these effects and than to predict
the $\ga \ga$ cross section with high enough accuracy.

Let us introduce a quantitative characteristic of the effect of the
photon-hadron scaling violation. According to the VDM
\beq
\si_{\ga p}(s)=\sum\limits_\scV  P_{\ga\to \scV }(s) \si_{\scV p}(s),
\label{vdm}
\eeq
where $P_{\ga\to V}(s)$ is the probability of photon  conversion to a
vector meson $V$; $\si_{\scV  p}(s)$ is the total cross section of $Vp$
interaction. Making use the AQM  \cite{akm} and U(3) symmetry of
meson-nucleon interactions gives
\beq
\si_{\scV  p}(s)= \frac23 \bar \si_{\scN p} \,(3s/2), \qquad
V=\rho,\omega,\phi.
\label{vp}
\eeq
This allows to separate in (\ref{vdm}) the contributions
of $\rho$, $\omega$, $\phi$ mesons:
\beq \si_{\ga p}^{(0)}(s)= P_{\ga\to\rho\omega\phi}(s)
               \frac23 \, \bar \si_{\scN p} \, (3s/2)
\label{si_gap_0}
\eeq
where
\beq
P_{\ga\to\rho\omega\phi} (s) =\sum \limits_{V=\rho,\omega,\phi}
              P_{\ga\to \scV} (s),
\qquad \bar \si_{\scN p}=\frac14 ( \si_{pp} +\si_{\bar p p}
+\si_{n p}
+\si_{\bar n p}).
\label{ga V}
\eeq
The argument in the right--handed part of Eq. (\ref{vp}) is
multiplicatively transformed ($3s/2$ instead of $s$), because the energies
of interacting quark pairs in $Vp$ and $Np$ systems are different. The
photon hadronization probability $P_{\ga\to\rho\omega\phi}\approx
P_{\ga\to\rho\omega\phi}^{(0)}= 1/250$ is known from the data on $\ga\to
V$ transitions on the mass shells of vector mesons. So $\rho,\omega,\phi$
contributions (i.e. $\si_{\ga p}^{(0)}(s)$ function) can be expressed only
in terms of the data on cross sections of nucleon-proton interactions. The
function $\si_{\ga p}^{(0)}(s)$, by virtue the specifying it experimental
data are independent and the physical sense is clear, is referred to as
the calibration curve for the cross section of $\ga p$ interaction. It is
clear, the deviation of measured $\si_{\ga p}(s)$ from the calibration
curve contains an additional information about photon hadronization, which
is not contained in $\si_{\ga p}^{(0)}(s)$.

For constructing the calibration curve in the energy region
$\sqrt{s}=5.93\div 22.97$ GeV, we use data from \cite{denisov,carrol}. In
the region $\sqrt{s}=30.4\div 62.7$ GeV, the data of
\cite{amos85,carboni84} on $\si_{pp}(s)$ and $\si_{\bar p p}(s)$ are used
(it seems to be reasonable to assume, that at such energies
$\si_{pp}=\si_{np}$, $\si_{\bar p p}=\si_{\bar p n}$ with a good
accuracy); for energies $\sqrt{s}=200,546,900, 1800$ GeV we use the
$\bar p p$ data from \cite{200,546,900,1800}.  These data, subjected to the
scale transformation $3s_{\scN p}/2=s_{\ga p}\equiv s$, are fitted by the
calibration curve
\beq
\si_{\ga p}^{(0)}(s)
=C\ln^2\frac{s}{s_0}+A(1+\frac{\Lambda_{01}^2}{s+\Lambda_{02}^2})
\label{5}
\eeq
\[
C=0.5777 \mkb, \quad \sqrt{s_0}=2.198 \GeV, \quad A=95.65 \mkb,
\]
\[
\Lambda_{01}=2.774\GeV, \quad \Lambda_{02}=3.589\GeV .
\]

The data and the calibration curve mentioned above, and also various
experimental points on $\si_{\ga p}(s)$ (including those of BUST) and two
variants of their fit, corresponding to the two variants of the high
energy points choice, are presented in Fig.1. The upper solid line
represents the cross section $\si_{\ga p}(s)$ as if the photon
hadronizates only in the strongly bound $\pi^+\pi^-$ system; physical
bases for invoking of this curve are discussed in section~3.

\begin{figure}[htb]
\epsfxsize=\textwidth \epsfbox{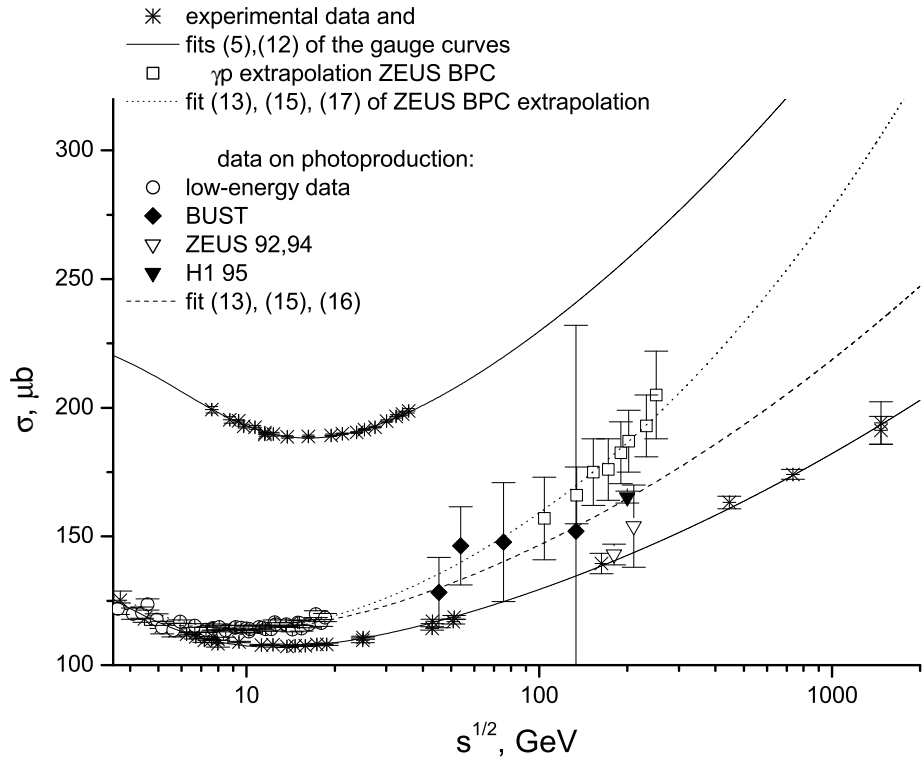}
\caption{Calibration curves and cross section of $\ga p$ interaction.}
\end{figure}

One can see from Fig.1, at low energies the photon-proton cross section
asymptotically approaches the calibration curve. This testifies, that
considerations used at constructing the calibration curve adequately
reflect the physics at these energies. Essential deviations from the
calibration curve emerge at $\sqrt{s}>8\GeV$. At $\sqrt{s}=200\GeV$ the
deviation of the cross section obtained by the photoproduction (direct)
method achieves $25\%$. That derived from the extrapolation \cite{ekstrap}
of ZEUS data \cite{DESYep} (indirect) is even larger. The BUST data,
obtained by the photoproduction method, are combined and fitted with both
direct and indirect measurements. The results are shown in Fig.1 as the
dashed and dotted lines correspondingly.

\section{Experimental technique and the results at the BUST.}

\subsection{The measured variables.
\label{2.1}}

In 1983 - 1989 at the BUST, the experiment on studying cross section
${d\sigma _{\mu A}^{(h)}(E_\mu, E_c, q^2)} / {dE_c}$ of inelastic
scattering of Cosmic Ray (CR) muons on atomic nuclei was carried out
\cite{novo89,n3,n4}.  Here $E_c = E_\mu-E_\mu ^\prime$ is the energy
transferred from muon into the hadronic cascade, $E_\mu $ and $E_\mu ^
\prime$ are muon energies in laboratory system before and after
interaction. The average number of nucleons in the nuclei of the
target was estimated as $\overline A=26$.
The events characterized by $E_c >700 \GeV$ were selected.

The average value of the  squared transferred four-momentum in our
experiment, $-\overline {q^2} = \overline {Q^2} = 0.5\ GeV^2$, is much
less than the characteristic squared energy  in the centre--of--mass system
of photon and nucleon, $s_{\ga \scN}>(40\ GeV)^2$ $(E_{\ga}\geq 700 \GeV)$.
This allows us to
pose the problem on experimental measurements of interaction cross section
$\sigma _ {\ga \scN} (s) $ of a real photon with a nucleon.

The problem of the measurement  $\si_{\gamma \scN}$ in the processes with
$\overline{Q^2}=0.5\GeV^2$ is nontrivial.  In  \cite{bb,bb2} it was shown,
that the differential cross section of inelastic $\mu A$ scattering at
small $Q^2$ can be represented as a product of $\si_{\gamma \scN}$ and a
certain function. This function depends on $Q^2$ and other parameters
taking into account the effects of shadowing in nuclei; its explicit form
is given in the Refs. mentioned above. Notice, the factorization of
$\si_{\gamma \scN}$  is a general property of $\mu A$ cross section at
small enough $Q^2$; the question is what is the particular value of $Q^2$
that the factorization can be actually performed at. In \cite{bb,bb2} it
is shown that this can be done if $Q^2 \le 10\GeV^2$. This
structure of $\mu A$ scattering cross section allows at
$\overline{Q^2} \ll 10\GeV^2$
to determine the $\si_{\gamma \scN}$ cross section, in spite of the fact
that  in each separate act of measurement the processes with $Q^2 \ne 0$
are fixed. In the expression for cross sections of inelastic $\mu A$
scattering (at $E_c > 700$ GeV inelastic $\mu A$ scattering initiates
a hadronic cascade), it is possible to factorize
the dependence from $Q^2$, to integrate over this dependence
and then to express the cross section directly through $\sigma _{\ga
\scN}(E_c)$  \cite{bb2}:
\beq
\frac {d\sigma _{\mu A}^{(h)}(E_\mu , E_c)}{dE_c} =
\frac {\alpha}{2\pi }\frac{AE_c}{E_\mu^2} \cdot
\sigma _{\gamma N}(E_c)\cdot F_A(E_\mu,E_c),
\label{bb10}
\eeq
where $F_A(E_\mu,E_c)$ is a function which takes into account
shadowing of nucleons in a nucleus.

In our experiment $\overline {Q^2}$ is determined as
\[
\overline {Q^2} = \frac{f_1(E_c)}{f_0(E_c)},
\quad f_n(E_c) =\int_{E_c}^{\infty} \frac {dN_\mu(E_\mu)}{dE_\mu} dE_\mu
\int_{Q_{min}^2}^{Q^2_{max}}
\frac{d^2 \si (E_\mu ,E_c,Q^2)}{dE_c dQ^2} \cdot Q^{2n} dQ^2,\quad n=0,1,
\]
where $N_\mu(E_\mu)$ is the integral energy spectrum of CR muons (see Eq.
(\ref{nmu}) ). At calculating the integral over $Q^2$, we use
the expression of $d^2 \si (E_\mu,E_c,Q^2)/dE_c dQ^2 $ and $\si _{\ga N}(E_c)$
from \cite{bb2} (this can be done because $\si _{\ga N}(E_c)$ describes
the experimental data fairly well) and restricted the integration range
from $Q^2_{min}$ to 10 GeV$^2$. The contribution of the range $Q^2 >$
10 Gev$^2$ in $\overline {Q^2}$ value is $<1\%$.

The integral energy spectrum of muons for zenith angles~$\theta<25^\circ$
was obtained in the special gauge experiment \cite{n4} and with an
accuracy of $5 \%$ approximated by the expression:
\beq
N_\mu (>E;\theta,\phi ) = k\cdot (E + E_{\small H})^{-2.83},
\label{nmu}
\eeq
where $E_{\small H}=230\GeV$, $k=const$. At $\theta>25^\circ$ we used
the calculated muon spectrum obtained in Ref. \cite{n5,n6}.

The BUST has sizes $16\times 16\times 11\ m^3$ and consists of
four horizontal and four vertical scintillation planes \cite{Alex79}.
In experiment \cite{novo89,n3,n4} the four horizontal scintillation
planes of the BUST together with the concrete layers between them were
used as a calorimeter that registered electromagnetic and hadronic cascades
generated by CR muons either in the rock immediately above the telescope
($ {\overline A} =25 $) or in the matter of the telescope
($\overline{A} =27$). Experimentally measured values are the
differential spectra of energy deposition in the calorimeter from the
electromagnetic ($j=e $) and hadronic ($j=h $) cascades
\beq
\frac{dN_{j}(\ve)}{d \ve} =
\int \frac {\dd N_\mu(E_\mu;\theta,\phi)}{\dd E_\mu} \cdot
\frac{d \sigma^{(j)}_{\mu A}(E_\mu,E_c)}{d E_c} \cdot n \cdot
W(E_c,\ve ;\theta ,\phi) S(\theta,\phi) dE_\mu dE_c \sin\theta
d\theta d\phi.
\label{nn4}
\eeq
Here $\ve$ is the sum of energy depositions in four horizontal
scintillation layers of the BUST; \\ $\dd N_\mu (E_\mu; \theta, \phi)/\dd
E_\mu $ is the differential spectrum of muons in a given direction at the
BUST location; \\
$d \sigma^{(j)}_{\mu A} (E_\mu, E_c)/d E_c $ is the production cross section of
electromagnetic or hadronic cascade with energy $E_c$ by muon with energy $E_\mu $; \\
$n$ is the number of nuclei in a unit volume of the detector (or of the rock); \\
$W(E_c,\ve; \theta, \phi)$ is the probability of that a cascade of
energy $E_c $ deposits the energy
$\ve $ in the telescope; \\
$S(\theta, \phi)$ is  the area of the telescope in a direction $\theta,\phi$.

At calculating the probability $W(E_c, \ve; \theta, \phi)$,
electromagnetic and hadronic cascade curves, obtained by Monte
Carlo simulation in view of real structure of the facility,
are used. $W(E_c, \ve; \theta, \phi)$ takes into account
the integration over the thickness of the target, therefore only
the integration over $S(\theta ,\phi)$ remains in Eq. (\ref{nn4}).

    Muons produce cascades in the following processes:\\
a) bremsstrahlung of photons, \\
b) production of $e^+e^-$ - pairs, \\
c) production of high-energy knock-on electrons, \\
d) inelastic scattering on nuclei.\\
The hadronic cascades are produced only in the process (d),
the processes (a), (b), (c) produce the electromagnetic cascades.
For the cross sections of the processes (a), (b), (c) the expressions
were taken from Refs. \cite{n7,bb3}.

To obtain the cross section $\sigma _{\ga \scN}(E_c)$ we use the
ratio of numbers of hadronic and electromagnetic cascades
\beq
R(\Delta \ve )= \frac {N_h(\Delta \ve)} {N_e(\Delta
\ve )},
\label{nn7}
\eeq
having assumed the cross sections of electromagnetic processes
are known. Making use of the ratio $R(\Delta \ve)$ has an important
advantage in comparison with the procedure of $\si _{\ga N}$
determination directly from the data on hadronic cascades.
The possible uncertainties in determination of $N_\mu (>E;\theta,\phi)$,
$W(E_c, \ve; \theta, \phi)$, $S(\theta, \phi)$ (see Eq.(\ref{nn4}))
are nearly completely disappear in the ratio (\ref{nn7}).

In an individual event we do not try to restore
the cascade energy $E_c$ on the energy deposition $\ve$, because
the error is equal to $30-60\%$. However, with a good enough
accuracy of $2-3 \%$,
it is possible to calculate the average energies of electromagnetic
and hadronic cascades creating the energy deposition in the facility
in a given interval $\ve _i \leq \ve < \ve _{i+1}$.

The properties of the BUST allow us to distinguish cascades generated on
different muon trajectories (in the cases when the cascade is accompanied
by a muon group), and two consecutive cascades on one and the same
trajectory and, thus, to exclude errors in the determination of $\ve$
caused by these factors.

\subsection{Separation of cascades. \label{2.2}}

To separate hadronic and electromagnetic cascades, we used
the number of $\pi \to \mu \to e$ decays recorded in the event
\cite{novo89,n3}. In hadronic cascades the number of $\pi - \mu - e$
decays is 20-25 times more, than in electromagnetic ones.
Recording the delayed pulse on the screen of 10-beams oscillograph
is the feature of $\mu - e$ decay (each beam corresponds to the
plane of the BUST). The separation procedure is based on the results of
electromagnetic cascades modeling and consists in the following.

1) The probability $P(m,\ve)$ of recording of $m$ decays in the
electromagnetic cascade, which creates the energy deposition $\ve$ in the
facility, is calculated \cite{n3}.

2) The separation criterion $m_0(\ve)$ is determined, such that
the probability $\chi$ of recording the number
of decays $m > m_0(\ve)$ in the
electromagnetic cascade is $\le 10^{-2}$,
\beq
\sum _{m=0}^{m_0} P(m;\ve) = 1 - \chi (m_0,\ve) \simeq
0.99.
\label{10}
\eeq

3) Cascades with $m \le m_0(\ve)$ are considered as electromagnetic ones.

4) The calculated number of electromagnetic cascades with
 $m > m_0(\ve)$ is subtracted
 from the number of cascades with $m > m_0(\ve)$ (i.e. it is taken
 into account that the "tails" of distributions $P(m,\ve )$
 fall into the area $m > m_0(\ve) $).

On the basis of (\ref {10}) one can write
\[
N(m\le m_0(\ve)) = N_e(\ve) - N_e(\ve)\cdot \chi
(m_0,\ve),
\]
where $N_e(\ve)$ is the actual  number of electromagnetic cascades.
Therefore the total number of electromagnetic cascades in the
interval $\Delta \ve$ is
\[
N_e(\Delta \ve) = {{N(m\le m_0(\Delta \ve))} \over
{1-\chi(m_0,\Delta \ve)}}.
\]
The remaining cascades are considered as hadronic ones.

5) We do not take into account hadronic cascades falling into the area of
electromagnetic ones. The calculations show, that the fraction of hadronic
cascades with $m \le m_0(\ve)$ does not exceed 3 $\%$. It is much less
than the statistical error, which is estimated as $\sqrt{N_h}$.

Thus, the efficiency of the electromagnetic cascades separation from the
hadronic ones is $ \ge~99 \%$; in this case the efficiency of separation
of hadronic cascades from electromagnetic ones is $\simeq~97 \%$.

\subsection{The results of measurements.\label{2.3}}

In Table 1 the data are presented on the numbers of electromagnetic $N_e
(\Delta \ve) $ and hadronic $N_h(\Delta \ve)$ cascades recorded in various
intervals of energy deposition in the facility $\Delta \ve = \ve _{min}
\div \ve _{max}$. The last interval of $\Delta \ve$ is not limited from
the above: $\ve > 573\ GeV$. The spectra are nonmonotone, because at small
$\ve$ the observation time was smaller (the total time was $T_{rec}$
= 31650 hours).

The average energy $\overline E_c^{(h)}$ of hadronic cascades, creating
energy deposition in an interval $\ve _{min} \le \ve < \ve _{max}$, is
identified with the energy of photon, $\overline E_\ga \equiv \overline
E_c^{(h)}$. For a given $\Delta \ve$, the photon energy interval $E_{\ga
(min)}\div E_{\ga (max)}$ is calculated. The last two columns of Table 1
comprise the photon--proton energies in the centre--of--mass system and
the corresponding total cross sections; energies being given in GeV, cross
sections being given in $\mkb$.

\begin{table}[htb]
\caption{The data on cascades and cross sections of $\ga N$ interaction}
\[
\begin{array}{ccccccccc}
\hline \ve_{min} &  \ve_{max} & N_e(\Delta\ve) &
N_h(\Delta\ve) & E_{\ga(min)} &  E_{\ga(max)} &
\overline{E_{\ga}} & \sqrt{s} & \si_{\ga N}
\\
\hline
77  & 133 & 1050 & 96  & 744  & 1260 & 1100 & 45.4 & 128 \pm 14
\\
133 & 230 & 1006 & 102 & 1260 & 2100 & 1550 & 53.9 & 146 \pm 15
\\
 230 & 573 & 445 & 45 & 2100 & 4980 & 3030 & 75.4 & 148 \pm 23
\\
573 &     &  36  & 4   & 4980 &      & 9500 &  133.5 & 152 \pm 80
\\
\hline
\end{array}
\]
\end{table}

The information about $\sigma_{\gamma N}(s)$ is extracted from
the ratio of numbers of hadronic and electromagnetic cascades, as
it has mentioned in section 2.1 (Eq. (\ref{nn7})).
The final results for $\si _{\ga N}(E_\ga)$ are obtained by integration of
photon-nucleon interaction cross section over the energy interval $\Delta
E_\ga=f(\Delta\ve)$ in view of the cross section dependence on energy. For
this purpose the forecast of $\si _{\ga N}(E_\ga)$ offered in \cite{bb2}
and consecutively refining fit of experimental data are used, on which the
calculated value $R_{cal}(\Delta \ve_i)$ is determined. The deviation of
experimental value $R _{exp}(\Delta \ve_i)$ from the calculated one
determines the deviation of actually measured cross section from the
forecast.

Now we can compare the method of $\si _{\ga N}$ measurements used in this
experiment with those used in DESY experiments.

1. Direct method \cite{zeus94,aid95} is based on photoproduction
registration. The photon being on-mass-shell (the virtuality is
negligible) is ensured by strict selection of the final hadronic states.

2. In the indirect method \cite{DESYep,ekstrap} electron in the inclusive
process $ep \to e + X$ is registered at small $Q^2$; the value of $\si
_{\ga N}$ is obtained by model--dependent extrapolation of the structure
functions to $Q^2=0$.

3. In the method described in this paper the photoproduction of
slightly virtual photon is registered. Since the cross section
falls down rapidly at
increasing $Q^2$, the absolute majority of the events occur at very low
$Q^2$, $\overline{Q^2}=0.5\GeV^2$. At such small $Q^2$, the cross section
$\si _{\ga N}$ is theoretically proved to be a multiplier at inelastic
$\mu A$ scattering sross section (Eq. (\ref{bb10})).

\section{About possible physical reasons of rise of $\ga N$ interaction
cross section.}

Two problems arise on a way to reduce photon--hadron interactions to
hadron--hadron ones. The first problem concerns the initial hadronic
states. In $\ga p$ interaction these states are a proton with the known
quark--gluon structure and a quark bag, created by a photon. The structure
of the latter is unknown and, moreover, statistical by its nature. In the
framework of the AQM the problem is formulated as follows: what number of
additive quarks describes the quark bag created by the photon? This
question naturally arises, when one takes into account big widths of
vector resonances. At transition of a photon into $\rho-$meson, two
alternative situations are possible: either $\rho-$meson is considered as
an initial hadronic state and interacts with the proton as a single
particle, or the initial state is the strongly bound $\pi^+\pi^-$ system
having originated in the decay of $\rho-$meson \cite{bb}. One of the
possible representations of this $\pi^+\pi^-$ system is a bag consisting
of four additive quarks. In more general case the average number of
additive quark-antiquark pairs must be considered as a function of energy
of $\ga p$ interaction.

The second problem is the dependence of photon hadronization probability
$P_{\ga \to V} (s)$ on the energy. At $\sqrt{s}>10\GeV$ it can be higher
than the standard value $1/250$ on $10 - 20 \%$, for example, due to the
nondiagonal transitions $\ga \to V^* \to V$, where $V^*$ is a
highly--excited resonance, $V$ is a basic
vector state ($V=\rho,\omega,\phi ...$) \cite{bb}.

Let us turn to the phenomenological model that takes into account the
above mentioned effects. Assume, that formation of the strongly bound
$\pi^+\pi^-$ system in the act of $\ga\to V\to 2\pi$ transition and
its subsequent interaction with the proton
is described by the model of four additive quarks. In this model the cross
section of $2\pi \, p$ interaction is estimated by the formula \\
\[
\si_{2\pi p}(s)=2\overline \si_{\pi p}(s/2) \qquad \overline\si_{\pi
p}=\frac12(\si_{\pi^+ p}+\si_{\pi ^-p}).
\]
Let $P_{\scV\to 2\pi}(s)$ be the
probability of conversion of a vector meson $V$ into $2\pi$-system before
interaction of $V$ with the proton. Then the cross section of $\ga p$
interaction can be written as
\beq
\begin{array}{l} \di
\si_{\ga p}(s)=\sum\limits_{\scV} P_{\ga \to \scV}(s)
\left(1-P_{\scV\to 2\pi}(s)\right)\si_{\scV p}(s)
+\sum\limits_{\scV} P_{\ga \to \scV}(s) P_{\scV\to 2\pi}(s)
\si_{2\pi p}(s)
\\[5mm]  \di
=\frac23\sum\limits_{\scV} P_{\ga \to \scV}(s)
\left(1-P_{\scV\to 2\pi}(s)\right)\overline \si_{\overline N p}(3s/2)
+2\sum\limits_{\scV} P_{\ga \to \scV}(s) P_{\scV\to 2\pi}(s)
\overline \si_{\pi p}(s/2)
\end{array}
\label{sigap}
\eeq
Formally substituting  in (\ref {sigap}) $ \sum\limits _ {\scV} P _ {\ga\to
\scV} =P _ {\ga\to\scV} ^ {(0)} =1/250 $, $P _ {\scV\to 2\pi} =1 $, we
obtain the so called conditional calibration curve
\[
\si_{\ga p}^{(1)} (s)=2P_{\ga\to\scV}^{(0)} \overline
\si_{\pi p}(s/2)
\]
For its constructing we use data of groups \cite{denisov,carrol}. The fit
of these data has resulted in the following expression
\beq
\si_{\ga p}^{(1)}
=C_1\ln^2\frac{s}{s_1}+A_1
\left(1+\frac{\Lambda_{11}^2}{s+\Lambda_{12}^2}\right)
\label{cal1}
\eeq
\[
C_1=1.155\mkb, \quad \sqrt{s_1}=1.672\GeV, \quad A_1=152.0\mkb,
\]
\[
\Lambda_{11}=5.006\GeV, \quad \Lambda_{12}=6.658\GeV.
\]
Due to the universality of the Froissart asymptotic of hadronic cross sections,
the parameter $C_1 = 2C$ is fixed by combinatorial reasons.  In Fig.1  the
curve (\ref {cal1}) is represented by the upper solid line.

Rewriting (\ref {sigap}) in terms of calibration curves reads
\beq
\si_{\ga p}(s) =k(s)\left[ \left( 1-\overline P_{\scV\to
2\pi}(s)\right) \si_{\ga p}^{(0)}(s)
+ \overline P_{\scV\to 2\pi}(s) \si_{\ga p}^{(1)}(s) \right],
\label{gap_cal01}
\eeq
where
\beq
k(s)=1+\frac1{P_{\ga \to V}^{(0)}}\sum\limits_{V,V^*} \Delta
P_{\ga\to V^* \to V}(s),
\label{kots}
\eeq
\[
\Delta P_{\ga\to V^* \to V}(s) = \sum\limits_{V,V^*} P_{\ga\to V^* \to
V}(s) - P_{\ga\to\scV}^{(0)},\ \ P_{\ga\to\scV}^{(0)} \simeq \sum
\limits_{V=\rho,\omega,\phi}
              P_{\ga\to \scV} (s)
\]
(see Eqs. (\ref {si_gap_0}), (\ref {ga V})).

The function $k(s)$ takes into account the rise of photon
hadronization probability due to nondiagonal transitions;
$\overline P_{\scV\to 2\pi}$ is the pionization probability averaged
over all vector mesons.

One can easily see from Fig.1, that all measured cross sections of $\ga p$
interactions lie between the two calibration curves. This allows to
assume, that the data can be interpreted within the framework of a
hypothesis about the certain dependence of the pionization probability on
energy or, what is the same, of the emergency probability quark bag
consisting of four additive quarks.

If we neglect all other factors increasing a photon hadronization probability
(in particular, the nondiagonal transitions) and  put in (\ref {gap_cal01})
$k(s)=1$, then the probability $\overline P_{\scV \to 2\pi}(s)$
can be expressed only through the experimental data:
\[
\overline P_{\scV \to 2\pi}(s)=\frac{\si_{\ga p}-\si_{\ga
p}^{(0)}}{\si_{\ga p}^{(1)}-\si_{\ga p}^{(0)}}
\]
The values, extracted in such way and their fits by the formula
\beq
\overline P_{\scV \to 2\pi}(s)= w \ln^n \frac{s}{s_2}
\label{probab}
\eeq
are shown in Fig. 2.

\begin{figure}[htb]
\epsfxsize=\textwidth \epsfbox{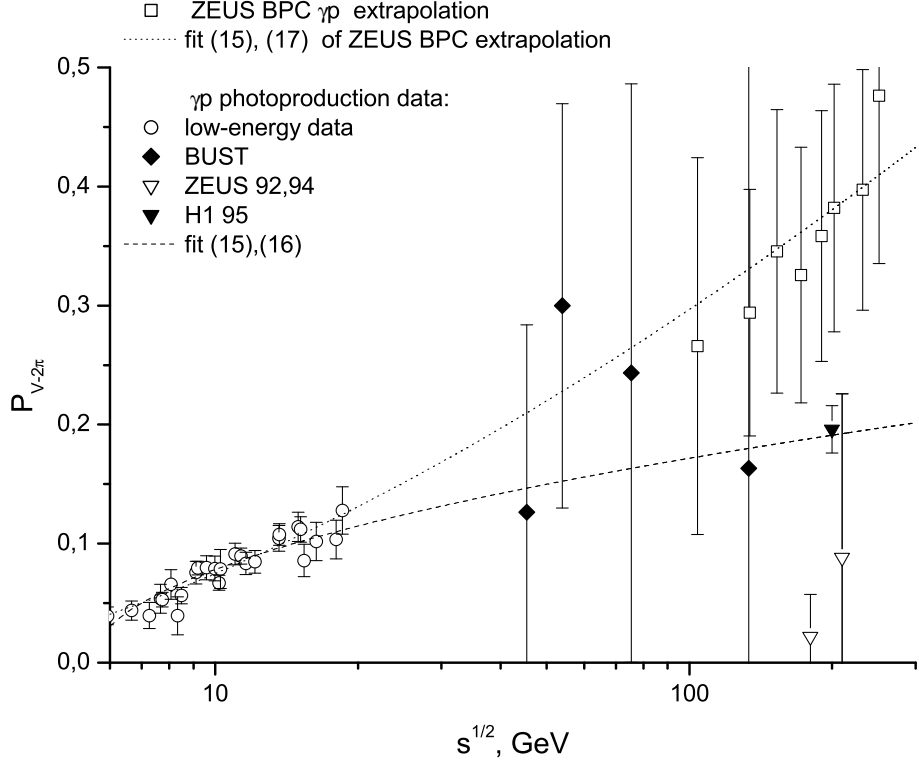}
\caption{Probability of vector meson pionization in
               $\ga p$ interaction.}
\end{figure}

The first fit,
\beq
w=0.0712, \; \sqrt{s_2}=5.465\ GeV,  \; n=1/2 \qquad \mbox{ ---
for direct measurements;}
\label{probab-dir}
\eeq
represented by the dashed line, is based on
the BUST data and the direct data \cite{zeus94,aid95}.
The second fit,
\beq
w=0.0140, \; \sqrt{s_2}=2.166\ GeV,  \; n=3/2 \qquad \mbox{ ---
for indirect data;}
\label{probab-indir}
\eeq
based on the BUST data and the indirect data
\cite{ekstrap}, is shown by the dotted line.

Substituting (\ref {probab}) in (\ref {gap_cal01}), one derives the two
curves for $\si _{\ga p}(s)$ fitting the two sets of experimental data
mentioned above, which are presented in Fig. 1.

The results obtained, in our opinion, testify that the average number of
additive quarks in the products of photon hadronization monotonously
increases with energy under the law
\[
\overline n_{\ga\to h}(s)=2\left(1+w\ln^n\frac{s}{s_2}\right)
\]
To obtain additional information, let us turn to the
results of modeling of $\ga \ga$ interaction cross section on
the basis of the $\ga p$ one.

\section{Modeling of $\ga \ga$ interaction cross section
and its comparison with the experimental data.}

The cross section of $\ga \ga$ interaction is expressed through the same
probabilities of photon hadronization and cross sections of hadronic
interactions, which appear in the model of $\ga p$ interactions in
Section~3. The first step of the theoretical cross section calculation is
obvious:
\beq
\begin{array}{l} \di
\si_{\ga\ga}(s)=k^2(s)\left(P_{\ga\to\scV}^{(0)}\right)^2
\left[\left(1-\overline P_{\scV \to 2\pi}(s)\right)^2\si_{\scV\scV}(s)
\right.
\\[5mm] \hspace*{10mm} \di
\left.
+2\overline P_{\scV \to 2\pi}(s)\left(1-\overline P_{\scV \to 2\pi}(s)\right)
               \si_{2\pi \scV}(s)
+\overline P_{\scV \to 2\pi}^2(s)\si_{2\pi2\pi}(s) \right].
\end{array}
\eeq
Further on we express the cross sections of meson and two--pion interactions
through the experimentally measured ones, simultaneously performing
(according to the AQM) the scale transformation of arguments:
\beq
\begin{array}{l} \di
\si_{\ga\ga}(s)=k^2(s)\left(P_{\ga\to\scV}^{(0)}\right)^2
\Biggr[\left(1-\overline P_{\scV \to 2\pi}(s)\right)^2 \cdot
\frac49\, \si_{\scN p}(9s/4)
\\[5mm] \hspace*{10mm} \di
+2\overline P_{\scV \to 2\pi}(s)
\left(1-\overline P_{\scV \to 2\pi}(s)\right)
  \cdot   \frac43 \, \si_{\pi p}(3s/4)
+\overline P_{\scV \to 2\pi}^2(s) \cdot
4 \frac{\si_{\pi p}^2(3s/8)}{\si_{\scN p}(9s/16)} \Biggr].
\end{array}
\label{gaga}
\eeq
At obtaining (\ref {gaga}) we use one more relation of the AQM:
$\si_{\pi\pi}(s)={\si_{\pi p}^2(3s/2)}/{\si_{\scN p}(9s/4)}$.
The final result is convenient to present via
the calibration curves:
\beq
\begin{array}{l} \di
\si_{\ga\ga}(s)=k^2(s)\cdot \frac{1}{250}\cdot \frac 23 \Biggl[
\left(1-\overline P_{\scV \to 2\pi}(s)\right)^2 \cdot \si_{\ga
p}^{(0)}(3s/2)
\\[5mm] \hspace*{10mm} \di
+2\overline P_{\scV \to 2\pi}(s)
\left(1-\overline P_{\scV \to 2\pi}(s)\right)
  \cdot \, \si_{\ga p}^{(1)}(3s/2)
+\overline P_{\scV \to 2\pi}^2(s) \cdot
\frac{\left(\si_{\ga p}^{(1)}(3s/4)\right)^2}%
{\si_{\ga p}^{(0)}(3s/8)} \Biggr].
\end{array}
\label{prognoz}
\eeq

In Fig. 3 the experimental data on $\si_{\ga\ga}(s)$ cross
section from \cite{pdg}  and the theoretical predictions
(\ref {prognoz}) at
$k(s)=1$ for two variants of photon hadronization probability
(\ref {probab-dir})
and (\ref {probab-indir})  are shown.

\begin{figure}[htb]
\epsfxsize=\textwidth \epsfbox{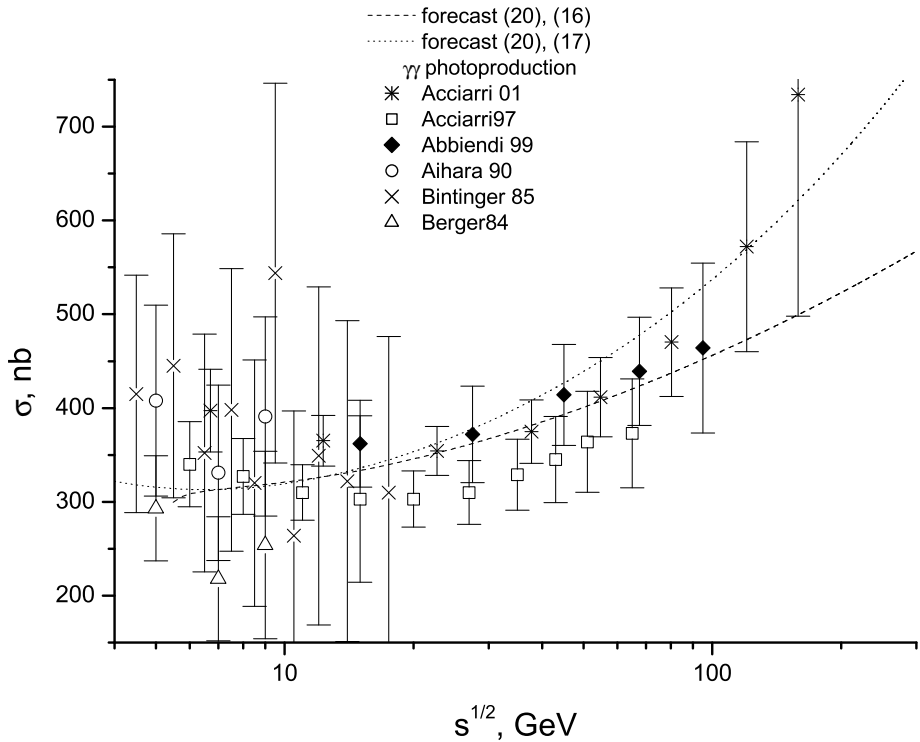}
\caption{Cross section of $\ga \ga$ interaction and its modeling.}
\end{figure}

The experimental uncertainties for this reaction, as well as those for
$\ga p$ one, do not allow to fix unambiguously the variant of high-energy
behaviour of $\si _{\ga p}(s)$ and $\si _{\ga \ga}(s)$. However, the data
on $\si _{\ga \ga}(s)$ at $\sqrt{s} < 100\ GeV$, which have rather small
errors, agree better with the forecast made on the basis of direct
measurements of $\ga p$ cross sections.

We would like to thank Dr. V. Bakatanov for the help in carrying out the
experiment, Prof. L. Bezrukov, Prof. S. Mikheyev and Dr. A. Butkevich for
the helpful discussions. This work is supported by the Federal Program
"Integration", Grant No. E0157.

\end{document}